\documentclass[12pt,nofootinbib]{revtex4-2}
\usepackage{amsmath,amssymb}
\usepackage{physics}
\usepackage{graphicx}
\usepackage{bm}
\usepackage{hyperref}
\usepackage{tikz}
\usetikzlibrary{patterns}

\begin{document}

\title{Is There Only One Vacuum?\\
{\small Asymptotic Boundary, Vacuum Energy, and the Limit of QFT in dS Spacetime}}
\author{Xinmu Zong}
\email{25la020w@rikkyo.ac.jp}
\affiliation{Department of Physics, Rikkyo University, Toshima, Tokyo 171-8501, Japan}

\author{Chunyu Huang}
\email{chunyuhuang@nuaa.edu.cn}
\affiliation{Key Laboratory of Radar Imaging and Microwave Photonics, Nanjing University of Aeronautics and Astronautics, Nanjing, Jiangsu 211106, China}

\begin{abstract}
This work proposes a reconstruction of the quantum field theory (QFT) scattering framework: the path integral governs an interaction kernel region, while the Hilbert space encodes asymptotic free boundary conditions. We critically reexamine foundational assumptions and standard observables of QFT, raising the possibility of a unique vacuum. We also introduce a new thermodynamic perspective on vacuum loop contributions. The motivation behind certain vacuum energy cancellation schemes is reconsidered, and the applicability of QFT to spacetimes without asymptotic flatness, such as de Sitter (dS) space, is questioned at a fundamental level.
\end{abstract}

\maketitle

\section*{The framework of QFT scattering}

In QFT, scattering can be calculated under the following framework: a whole spacetime is divided into two parts: A kernel bulk where interaction may happen, with the path integral traveling all the way through it, and, an asymptotic boundary where a free theory can apply, with the Hilbert space defined there as a boundary condition.

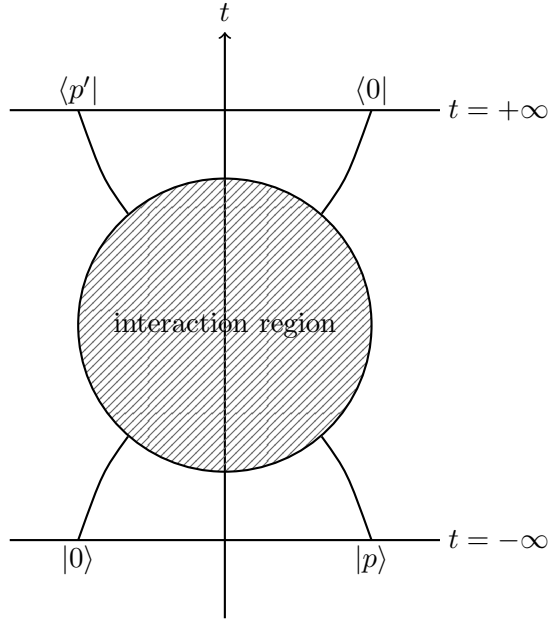
\begin{figure}[htbp]
\centering
\begin{tikzpicture}[scale=1.3]

  \draw[->, thick] (0, -3) -- (0, 3) node[above] {$t$};

  \draw[thick] (-2.2, 2.2) -- (2.2, 2.2);
  \draw[thick] (-2.2, -2.2) -- (2.2, -2.2);

  \node at (2.8, 2.2) {$t = +\infty$};
  \node at (2.8, -2.2) {$t = -\infty$};

  \node at (-1.5, -2.4) {$\ket{0}$};
  \node at (1.5, -2.4) {$\ket{p}$};
  \node at (-1.5, 2.4) {$\bra{p'}$};
  \node at (1.5, 2.4) {$\bra{0}$};

  \draw[thick] (-1.5, -2.2) .. controls (-1.25, -1.5) .. (-0.99, -1.14);
  \draw[thick] (1.5, -2.2) .. controls (1.25, -1.5) .. (0.99, -1.14);
  \draw[thick] (-0.99, 1.14) .. controls (-1.25, 1.5) .. (-1.5, 2.2);
  \draw[thick] (0.99, 1.14) .. controls (1.25, 1.5) .. (1.5, 2.2);

  \fill[pattern=north east lines, pattern color=black!60] (0,0) circle(1.5);
  \draw[thick] (0,0) circle(1.5);
  \node at (0, 0) {interaction region};

\end{tikzpicture}
\caption{A hybrid framework.}
\label{fig:hybrid}
\end{figure}

Instead of defining a Hilbert space in the interaction region and then constructing interaction picture's states and operators, take the path integral as an axiom which describes time evolution whether or not involving interaction \footnote{In interaction theories like QCD, the Hilbert space may even be ill-defined, while the path integral works well.}. The practice of placing the path integral in a more fundamental position when calculating interaction processes is also consistent with Haag's theorem \cite{Haag1955}, which challenges the validity of the interaction picture in quantum field theory. The Hilbert space, however, still plays a role in the asymptotic free region \footnote{Here, asymptotic free refers to the condition that interactions vanish at spatial or temporal infinity, not to the asymptotic freedom in QCD.}; there, particle states are defined through the limiting process as boundary conditions. Think of how LSZ reduction is deduced \cite{Srednicki}.

The LSZ reduction formula

\begin{equation}
\frac{\langle p' \dots | p \dots \rangle}{\langle 0|0 \rangle} =
\int i(2\pi)^{-3/2} e^{ipx}(-\partial^2 + m^2) \dots 
\int' i(2\pi)^{-3/2} e^{-ip'x'}(-\partial'^2 + m^2) \dots 
\frac{\langle 0 | T \phi' \dots \phi \dots | 0 \rangle}{\langle 0|0 \rangle} \tag{1}
\end{equation}     

with the $n$-point Green function / correlation function ($n \geq 0$)
\begin{equation}
\frac{\langle 0 | T \phi(x_1) \dots \phi(x_n) | 0 \rangle}{\langle 0|0 \rangle}
= \frac{\int [D\phi]\, \phi(x_1) \dots \phi(x_n) e^{iS}}{\int [D\phi]\, e^{iS}} \tag{2}
\end{equation}

build a bridge between path integral language and Hilbert space language, bulk integral behavior and boundary condition behavior, interaction involving theory and asymptotic free theory.

\section*{The role of vacuum amplitude}

LSZ reduction has some precondition including that the ground state is unique and separated from excited states, so $\langle 0|0 \rangle=1$, which can be regarded as a normalization condition. Instead of setting $\langle 0|0 \rangle=1$, one can always divide a $\langle 0|0 \rangle$ to normalize like what is done above. And one can write $\langle 0|0 \rangle = \int [D\phi]\, e^{iS}$ (defined up to a factor).

This path integral is divergent even in free theory. Think about a $k$-space loop $\circ$, or its exponential, $\exp(\circ)$, a loop is divergent, so $\int [D\phi]\, e^{iS}$ is divergent. If the factor is nonzero, $\langle 0|0 \rangle = \infty$; if the factor is 0 or say $\frac{1}{\infty}$, $\langle 0|0 \rangle$ can be any number between 0 and $\infty$.

In this sense, the value of $\langle 0|0 \rangle$ is not decided, and there is a physics reason. That is, $\langle 0|0 \rangle$ is not observable. One can imagine a scattering amplitude from incoming particles to outgoing particles, but who can imagine a scattering amplitude from vacuum to vacuum?

$\langle 0|0 \rangle$ is a condition, not an observable.

Up to this point, $\langle 0|0 \rangle$ can be any number. But in practical QFT calculations, one can pretend that $\langle 0|0 \rangle=1$ as long as one omits all the vacuum bubbles ($\exp(\circ)$ in free theory, $\exp(8\ldots)$ in $\phi^4$ theory).

In fact, $\langle 0|0 \rangle =1$ serves not only as a normalization condition, but also as a renormalization condition. Here are some optional renormalization conditions:

\begin{itemize}
    \item $\langle 0|0 \rangle = 1 \ \Longleftrightarrow$ omit vacuum bubble
    \item $\langle 0|\phi|0 \rangle = 0 \ \Longleftrightarrow$ omit tadpole, and not write linear counterterm $Y\phi$ in Lagrangian density $\mathcal{L}$
    \item $\langle p|\phi|0 \rangle = (2\pi)^{-3/2} e^{-ipx} \ \Longleftrightarrow$ OS scheme
\end{itemize}

and $\langle 0|0 \rangle=1$, $\langle 0|\phi|0 \rangle=0$, $\langle p|\phi|0 \rangle=(2\pi)^{-3/2} e^{-ipx}$ is exactly what one wants an asymptotic free vacuum state to act like when deducing LSZ reduction formula. Renormalization technique also includes conditions on vertex functions and coupling constants.

Although it is often said that renormalization has something to do with the fact that interaction alters the vacuum structure, one does not need such an interaction vacuum state formalism, just like one does not need Hilbert space language in an interaction region. One only needs the asymptotic free vacuum, $|0\rangle$ \footnote{If Spontaneous symmetry breaking effect exists, the vacuum will transition from unstable to stable but remain asymptotic free.}. Renormalization has everything to do with rescaling states and operators to connect the interaction region and asymptotic free region.

Despite the renormalization condition $\langle 0|0 \rangle=1$, in principle, $\langle 0|0 \rangle = \int [D\phi]\, e^{iS(\phi)}$ (defined up to a factor) can still be any number if choose other renormalization scheme. Instead of writing a specific factor, say $\langle 0|0 \rangle$ can be set arbitrarily.

\section*{Thermodynamics of vacuum}

Although unnormalized $\langle 0|0 \rangle$ is not an observable, it could have some deeper meaning other than a (re)normalization condition to be set.

Considering Euclidean path integral with zero temperature limit:

\begin{equation}
Z = \langle 0|0 \rangle = \int [D\phi]\, e^{iS} = \int [D\phi]\, e^{-S_E} \tag{3}
\end{equation}

and thermodynamics:

\begin{equation}
F = -\frac{1}{\beta} \ln Z, \qquad
F = \langle H \rangle - \frac{1}{\beta} S_{\text{TH}} \tag{4}
\end{equation}

We found an equation:

\begin{equation}
\langle H \rangle - \frac{1}{\beta} S_{\text{TH}} = -\frac{1}{\beta} \ln \langle 0|0 \rangle \qquad (\beta \to \infty) \tag{5}
\end{equation}

which seems to give (if $S_{\text{TH}} - \ln \langle 0|0 \rangle$ is not $\infty$):

\begin{equation}
\langle H \rangle = 0, \qquad S_{\text{TH}} = \ln \langle 0|0 \rangle \tag{6}
\end{equation}

If set $\langle 0|0 \rangle = \infty$, then thermal entropy $S_{\text{TH}} = \infty$. This means, if vacuum bubble is taken into account, they will contribute to vacuum entropy.

If set $\langle 0|0 \rangle = 1$, then thermal entropy $S_{\text{TH}} = 0 = S_{\text{EN}}$. This may seem like a trivial conclusion—that the vacuum has zero thermal entropy—and that the unitary condition $\langle 0|0 \rangle = 1$ corresponds to that the whole system has zero entanglement entropy $S_{\text{EN}} = 0$. However, this conclusion is not so trivial considering it still holds when interaction occurs somewhere. There is no information loss from an initial asymptotic free vacuum through an interaction zone to a final asymptotic in QFT framework.

It is interesting to notice that thermal entropy is related to the path integral, as shown in this equation and its deduction, and the fact that the direction of thermal entropy increase is the direction of time, and the path integral cannot reverse the time arrow.

Although notation $\phi$ is used, the deduction is based on path integral, so our equation is not limited to scalar field. The vacuum energy $\langle H \rangle$ comes to be zero, and we don’t see any boson and fermion energy cancellation here.

In conventional quantum field theory, vacuum energy is often said to receive divergent contributions from vacuum bubble diagrams, which motivates mechanisms such as supersymmetry to cancel bosonic and fermionic zero-point energies. However, according to the equation we derived above, the vacuum energy expectation value is zero regardless of whether vacuum loops are included or subtracted. The underlying divergence from vacuum loops contributes only as an additive constant to the thermal entropy, not to the vacuum energy itself.

\section*{Inclusion of gravity}

From now on, QFT in curved spacetime will be discussed.

It is also said that gravitation changes the structure of vacuum. However, this need not be the case. First, gravitation is not a force or an interaction in the usual sense. Gravitation is curved spacetime. Or think it another way, to feel the gravitational effect there must be some external force which fixes one in a point on the curved spacetime, or say manifold, preventing one from free-falling. Think of equivalence principle. Free-falling is equivalent to no gravitation and no external force. Free region is flat region. Second, as previously discussed, there could be just one vacuum, one asymptotic free vacuum in QFT. Last, such an external force or external source, typically requiring infinite energy, is responsible for the thermal bath.

The QFT framework applies in spacetimes that admit an asymptotically flat, or say free region, for example, the Schwarzschild black hole exterior. In fact, from black hole thermodynamics one has already known that no singularity condition poses a restriction so that Euclidean time $t_E$ has a period corresponding to inverse temperature $\beta$ in finite temperature field theory \cite{Hawking1977}. At finite $\beta$, $\langle H \rangle = \frac{1}{\beta}(S_{\text{TH}} - \ln \langle 0|0 \rangle)$, it is the first law of black hole thermodynamics with a log function. Same equation also appears in Hawking’s work \cite{Hawking1978}. The idea that the number of quantum states, or say the dimension of Hilbert space is given by $S_{\text{BH}}$ might be less a fundamental truth than an over-interpretation. In a framework where observables are defined only on asymptotically flat boundary, such assumptions become physically unfounded. Also, thermal entropy is not entanglement entropy. Although the detailed mechanism by which information is encoded in Hawking radiation after the Page time remains unknown, if the total entanglement entropy of the universe is zero, then no information can be lost in any scattering process, including gravitational ones. There is no black hole information paradox.

If the background spacetime has no asymptotically free or flat region, as in dS spacetime, QFT scattering framework does not apply from the very beginning \footnote{Proposals like the Bunch–Davies vacuum or the dS/CFT correspondence remain speculative frameworks without experimentally verifiable observables such as the S-matrix.} \footnote{One might wonder why AdS/CFT seems tractable while AdS is also curved everywhere. The key lies in the existence of a conformal boundary in AdS spacetime, which provides a stage for defining CFT observables.}. We do not think the vacuum energy in QFT has anything to do with the cosmology constant in GR, because the spacetime of the latter is not in the scope of the former. Also, in old-school string theory \cite{Polchinski}, a 2D CFT worldsheet embedded in a 10D target spacetime drives the vacuum Einstein equation $R_{\mu\nu} = 0$ via renormalization group analysis \footnote{We call this relation “RG flow = Ricci flow”, as a salute to ER = EPR.}. This derived vacuum does not naturally accommodate a cosmological constant.

Our argument does not hinge on whether QFT in dS spacetime or the dS/CFT correspondence exists. What we provide is a different line of reasoning: QFT (of which CFT is a special case) may fail to be well-defined in $\Lambda$CDM-type future universe (with dS space as its idealization) not due to observer dependence, not due to cosmological horizon and associated Gibbons-Hawking entropy, but simply because there is no asymptotically flatness, see also \cite{DeWitt}. This structural absence renders standard QFT observables, like the S-matrix, undefined. No flatness means no observable can be defined, making the description of dS-like quantum gravity through background-dependent frameworks even more problematic.

\section*{Conclusion and outlook}

This work reconstructs the QFT scattering framework by placing the path integral at kernel region and assigning the Hilbert space to asymptotic free or flat boundary. A thermodynamic relation involving vacuum loops is proposed, leading to a vanishing vacuum energy without invoking boson–fermion cancellation, thereby weakening one of the main physical motivations for supersymmetry. More fundamentally, it is argued that without asymptotic flatness, QFT loses its root, casting doubt on its applicability to spacetimes like dS space.

However, it is still too early to conclude that there is no quantum gravity theory which could and should describe dS universe. It is worth noting that the viability of QFT renormalization techniques is closely related to the existence of a global asymptotically flat boundary. Speculatively, if the gravity nonrenormalizable issue was resolved through ideas like asymptotic safety \cite{Percacci2011}, could this open a way toward meaningful observables even in the absence of asymptotic free vacuum and QFT observable?

\section*{Appendix: A tentative view on the motivation of quantum gravity} 

A particle in a superposition is a gravitational source in a superposition. Therefore, gravity must be quantized, even if it is non-renormalizable. This motivation is stronger than singularity (maybe the end of epistemology) or black hole information problem (no paradox). 

Eppley-Hannah’s thought experiment also argues for the necessity of quantizing gravity. It is about detecting a particle using a classical gravitational wave violates uncertainty principle or locality. 

Eppley-Hannah’s argument and the motivation of quantum gravity mentioned above share a same premise that a particle generates gravity. (Considering the analogy of classical electromagnetism, the matter-electromagnetic wave scattering is related to the forced vibrated charges generating new electromagnetic fields.)  

But what if particles are too light to generate gravity? This may sound weird, but not that weird considering Standard Model particles were massless without Higgs mechanics. The natural mass quanta is Planck mass. Particles with mass no smaller than one Planck mass is the dark matter which only couples with gravity because Standard Model breaks down at Plank scale. On the other hand, Standard Model particles are naturally particles with zero mass which are affected by gravity but do not generate gravity, like photon. For ordinary matter, most mass originates from the internal energy of Standard Model particle interactions. Gravity emerges only from energy exceeding the Planck threshold. 

More radically, gravity is fundamentally classical, separate from quantum. Gravity non-renormalization is not a problem to be fixed. Quantum gravity does not exist. 

Note: This appendix presents a speculative perspective and does not represent the core claim of this work. It serves only to explore the possibility that the motivation for quantizing gravity may not be as solid as often assumed.

\end{document}